\documentclass[twocolumn]{aastex631}

\usepackage{amsmath}
\usepackage{amssymb}
\usepackage{amsthm}

\usepackage{graphicx}
\usepackage{dcolumn}
\usepackage{bm}
\usepackage{subfigure}
\usepackage{url}
\usepackage{hyperref}
\usepackage{longtable}
\usepackage{natbib}
\usepackage{listings}
\usepackage[normalem]{ulem}
\usepackage{bm}
\usepackage{comment}
\usepackage{soul}

\usepackage{xspace}
\usepackage{xcolor, fontawesome}
\definecolor{twitterblue}{RGB}{64,153,255}

\newcommand{\twitter}[1]{\href{https://twitter.com/#1 }{\textcolor{twitterblue}{\faTwitter}\,\tt \textcolor{twitterblue}{@#1}}}

\newcommand{\github}[1]{\href{https://github.com/#1 }{\textcolor{black}{\faGithub}\,\tt \textcolor{black}{#1}}}

\newcommand{\stella}{\texttt{stella}}
\newcommand{\tess}{\textit{TESS}}

\usepackage{hyperref}

\submitjournal{ApJL}

\shorttitle{Self-Organized Criticality}
\shortauthors{Feinstein et al.}

\begin{document}

\title{Testing Self-Organized Criticality Across the Main Sequence \\
using Stellar Flares from \textit{TESS}}

\author[0000-0002-9464-8101]{Adina~D.~Feinstein}
  \altaffiliation{NSF Graduate Research Fellow}
  \affiliation{Department of Astronomy and Astrophysics, University of Chicago, Chicago, IL 60637, USA}
  
\author[0000-0002-0726-6480]{Darryl~Z.~Seligman}%
  \affiliation{Department of the Geophysical Sciences, University of Chicago, Chicago, IL 60637, USA}

\author[0000-0002-3164-9086]{Maximilian~N.~G{\"u}nther}
\altaffiliation{Juan Carlos Torres Fellow; ESA Research Fellow}
\affiliation{Department of Physics, and Kavli Institute for Astrophysics and Space Research, Massachusetts Institute of Technology (MIT), Cambridge, MA 02139, USA}%
\affiliation{European Space Agency (ESA), European Space Research and Technology Centre (ESTEC), Keplerlaan 1, 2201 AZ Noordwijk, The Netherlands}
  
\author{Fred~C.~Adams}
  \affiliation{Physics Department, University of Michigan, Ann Arbor, MI 48109}
  \affiliation{Astronomy Department, University of Michigan, Ann Arbor, MI 48109}

\correspondingauthor{Adina~D.~Feinstein;\\ \twitter{afeinstein20}; \github{afeinstein20};} \email{afeinstein@uchicago.edu} 

\begin{abstract}

Self-organized criticality describes a class of dynamical systems that maintain themselves in an attractor state with no intrinsic length or time scale. Fundamentally, this theoretical construct requires a mechanism for instability that may trigger additional instabilities \textit{locally} via  dissipative processes. This concept has been invoked to explain nonlinear dynamical phenomena such as  featureless energy spectra that have been observed empirically for earthquakes, avalanches, and solar flares. If this interpretation proves correct, it implies that the solar coronal magnetic field maintains itself in a critical state via a delicate balance between the dynamo-driven injection of magnetic energy and the release of that energy via flaring events. All-sky high-cadence surveys like the Transiting Exoplanet Survey Satellite (\tess) provide the necessary data to compare the energy distribution of flaring events in stars of different spectral types to that observed in the Sun. We identified $\sim 10^6$ flaring events on $\sim 10^5$ stars observed by \tess\ at 2-minute cadence. By fitting the flare frequency distribution for different mass bins, we find that all main sequence stars exhibit distributions of flaring events similar to that observed in the Sun, independent of their mass or age. This may suggest that stars universally maintain a critical state in their coronal topologies via magnetic reconnection events. If this interpretation proves correct, we may be able to infer properties of magnetic fields, interior structure, and dynamo mechanisms for stars that are otherwise unresolved point sources.

\end{abstract}


\section{Introduction}

The concept of self-organized criticality \citep{bak1988} describes a class of dissipative dynamical systems which remain at a critical point with no intrinsic length or time scale. The existence of the critical state requires a local instability, which occurs when some parameter exceeds its critical value and results in a dissipative transport process where this same parameter increases at neighboring sites. A simple physical analogy is a sand pile. As sand particles are added, the difference in height between neighboring sites on the pile increases. When the additional sand particles make the new height difference exceed a critical threshold, avalanche events occur. This system maintains a critical slope, representing a dynamical attractor that is insensitive to the initial conditions. This critical state is maintained via nonlinear avalanche events spanning all length scales triggered by perturbations. 

While the sand pile analogy is simplistic by construction, self-organized criticality naturally manifests in a variety of physical systems. Applications have been found in hydrodynamical turbulence, forest fires and other percolation systems \citep{Turcotte1999}, landslides \citep{Bak1990,Turcotte2002}, neuroscience \citep{Ribeiro2010,Hesse2014}, climate fluctuations \citep{Grieger1992}, rainfall \citep{Andrade1998}, accretion disks \citep{Dendy1998}, traffic jams \citep{Nagel1993}, evolution \citep{Bak1993}, extinction \citep{Newman1996}, financial markets \citep{Bak1997} and even Conway's game of Life \citep{bak1989}, to name a non-comprehensive list. The theory also explains the Gutenberg-Richter \citep{Gutenberg1956} law for the distribution of earthquake energies \citep{bak1989_earthquakes,Sornette1989,Olami1992}, 
\begin{equation}\label{eq:distribution}
  \frac{dN}{dE}\sim E^{-\alpha}\,,
\end{equation}
where $N$ is the number of earthquakes, $E$ is the energy released in the earthquake (where the earthquake magnitude $m\propto\log E$), and the power-law exponent $1.25<\alpha\sim1.5$. A 3-dimensional slip-stick model of tectonic events produces a critical exponent of $\alpha=1.35$ \citep{bak1989_earthquakes}, in reasonable agreement with the observed law. It is worth noting that scalar redistribution rules for the sandpile analogy have been generalized, and that the dynamics of the vectorial case are quantitatively similar to those for the scalar-field case \citep{Robinson1994}.

Even the solar coronal magnetic field may reside in a self-organized critical state \citep{lu1991}. This hypothesis naturally explains the power-law dependence of the magnitude of solar flares, which is of the same form as Equation (\ref{eq:distribution}), where $\alpha\approx1.4$ \citep{lu1991}. This characteristic exponent and the subsequent temporal clustering are universal between earthquakes and solar flares \citep{deArcangelis06}. 

Explosive events on the Sun are believed to be driven by the energy stored in twisted coronal magnetic field lines \citep{Parker1989}. Such field configurations are generated through dynamo action in the fluid-dominated interior \citep{Charbonneau2010}, and through convective and coriolis driven vortical subsurface plasma flows \citep{Parker1955,Moffat1978,Longcope1998,Seligman2014}. The concentration of magnetic field lines can lead to the release and subsequent dissipation of energy via the process of magnetic reconnection \citep{Sturrock1984}. These reconnection events could be triggered when the angle, $\theta$, between neighboring magnetic field vectors is greater than a critical value, $\theta_c$. The reconnection event changes the angles for neighboring field lines and can trigger additional events \citep{Sturrock1984,Parker1988,Sturrock1990}. When $\theta<\theta_c$, a sequence of metastable states develops, allowing for the buildup of non-potential magnetic energy in the form of twisted field lines. Thus, this configuration satisfies all the requirements for a self-organized critical system, if given a source for energy injection.  

Recent studies have begun to explore if flaring events on stars across all spectral types and evolutionary stages exhibit the same power-law distribution as in the Sun. This finding may suggest that other stars also maintain self-organized critical states in their coronal magnetic fields. However, this hypothesis has previously been difficult to test due to (i) the lack of a long observational baseline to capture a sufficient number of flaring events, (ii) the lack of observations of a large number of stars across spectral types, and (iii) the difficulty in observing, identifying and characterizing flares, especially those with low amplitudes relative to instrumental noise. Extra-solar surveys such as \textit{Kepler}/\textit{K2} \citep{Borucki10} and the Transiting Exoplanet Survey Satellite \citep[\tess;][]{ricker15} have provided solutions to (i) and (ii). These missions have provided unprecedented, high precision, long-baseline light curves for hundreds of thousands of stars. 

\citet{Aschwanden21} identified a power-law dependence with $\alpha=1.824\pm0.007$ in the cumulative distribution of flaring events observed in a set of stars across stellar types observed with \textit{Kepler}. This suggests that the self-organized critical state observed in the Solar corona is ubiquitous. However, the published \textit{Kepler} catalogs tend to be biased towards high-amplitude flares because of the inefficiency at identifying low-amplitude events of implemented flare-detection algorithms. Moreover, these catalogs are contaminated with other features easily mistaken for flaring events, such as rapidly rotating or variable stars \citep{Davenport16}. Further complicating matters, the flares that are both observed and identified correctly in \textit{Kepler} are not fully temporally resolved due to the 30-minute cadence of the observations. The \tess\ mission has recently provided 2-minute cadence light curves for $\sim$200,000 of the nearest and brightest stars. This data provides a full sample of spectral types across the main sequence to search for temporally-resolved flaring events \citep{guenther19_flares}.

Additionally, machine learning techniques curated to identify flares in \tess\ short cadence data have been developed to rectify issue (iii). These novel techniques are capable of identifying low-amplitude flares with high fidelity \citep{feinstein20,vida21}. This combination of unprecedented data and efficient identification techniques provides a new opportunity to expand the hypothesis of self-organized criticality in the solar corona to a sample of stars representative of the galactic census of spectral types and ages. In this letter, we provide a complementary search for indications of self-organized criticality in flaring events observed by \tess, using our newly created catalog of stellar flares from two years of data (G{\"u}nther et al, in prep).

\section{Observations with \tess}

NASA's  five-year \tess\ mission is currently performing time-series photometry of $\sim 90\%$ of the sky \citep{ricker15}. The survey observes in $24^\circ \times 96^\circ$ sectors for $\sim$\,27 days at a time. During its primary 2-year mission, \tess\ observed $\sim$200,000 pre-selected stars at 2-minute cadence across both ecliptic hemispheres. \tess\ has been providing an unprecedented data set at high temporal resolution with which to identify  flaring events across stellar spectral types and ages. The light curves used in this study were processed by the Science Processing Operations Center (SPOC) pipeline operated at the NASA Ames Research Center, which performs optimized aperture selection and systematics detrending \citep{Jenkins16}. We applied pipeline-assigned quality flags to mask contaminated\footnote{Contamination can originate from cosmic rays, reaction wheel desaturation events, and other spacecraft sources. For more information see Table 28 in \cite{tenenbaum2018}.} regions of the light curves.

\begin{figure}[t]
\begin{center}
\includegraphics[width=0.465\textwidth,trim={0.25cm 0 0 0}]{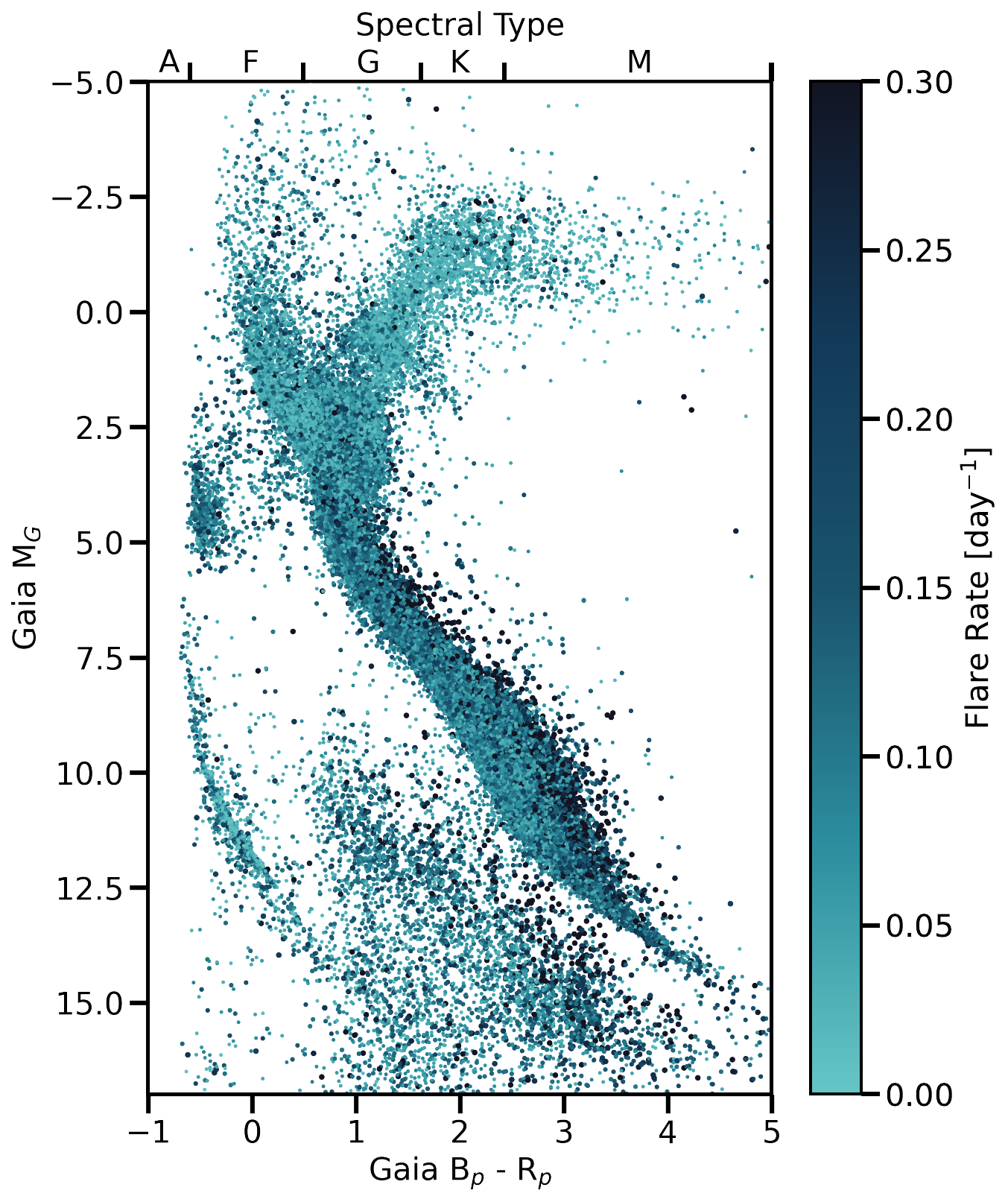}
\caption{A color-magnitude diagram Gaia $B_\textrm{p}-R_\textrm{p}$ color and absolute Gaia G magnitude, M$_{\textrm{G}}$) for our sample colored by flare rate. The flare rate was calculated by weighting each flare by its output ``probability" from \stella. Here, we accounted for all identified flares, not just flares with probabilities $\geq 0.9$. Stars towards the top of the main sequence tend to have higher flare rates. This trend could be due to being young and metal-rich or being binaries. \label{fig:hr}}
\end{center}
\end{figure}

\subsection{Flare Identification}\label{subsec:id}

As mentioned previously, flare identification in time series photometric data has proven challenging. Previous methods of flare identification have relied on detrended, or ``cleaned", light curves and have heuristics for outlier detections. However, detrending  often removes low-amplitude flares entirely, and the outlier-heuristics \citep[e.g][]{chang15} are biased towards the identification of only the highest amplitude events. Conversely, more lenient heuristics produce a significant number of false positive events.

Neural networks are a class of supervised machine learning algorithm optimized for visualization problems such as identifying features in 1-dimensional time series \citep[e.g.][]{ansdell18, pearson18, vida21} or images \citep[e.g.][]{tanoglidis21}. Here, we used the convolutional neural networks (CNN) developed in \cite{feinstein20}\footnote{The pre-trained CNNs are available online: \href{https://archive.stsci.edu/hlsp/stella}{https://archive.stsci.edu/hlsp/stella}.}, which are accompanied by the open-source software package \stella\ \citep{feinstein20_joss}, and are easily scalable to a large number of \tess\ 2-minute light curves. \stella\ is unique in that it provides a ``probability" that any given cadence in a light curve is or is not part of a flaring event. We ran all 2-minute \tess\ light curves through 10 pre-trained \stella\ models and averaged the probability outputs for our final catalog of flaring events. We require a probability $\geq 0.9$ for an event, meaning the event has a 90\% probability of being a true flare. We provide upper and lower limits  by only including flaring events with a probability $\geq 0.50$ and $\geq 0.99$, respectively.  

While \stella\ recovers a high percentage of real flaring events, it was trained on the relatively small data set presented in \cite{guenther19_flares}, with limited examples of contaminating features such as variable and eclipsing binary stars, and noisy light curves. In G{\"u}nther et al. (in prep) and this study, we therefore apply four quality control filters to the \stella\ outputs, mitigating the risk of false positives:

\begin{enumerate}
  \item \textit{Signal-to-Noise Ratio Filter}: This filter removes any false positives originating from the increased photon noise of faint targets. To do this, we estimated the root-mean-square (RMS) noise of a given light curve. We first remove any variability using a biweight filter with a window size of 20 minutes and then take the RMS of the flattened light curve. We require stellar flare candidates to have an amplitude of $\geq 3 \times$ RMS noise.
  \item \textit{Outliers}: This filter removes flaring events that appear in limited data points. We removed any flares with a fitted duration of $\leq 4$~minutes which correspond to 2 \tess\ data points.
  \item \textit{Eclipsing Binaries (EBs)}: This filter removes false positives originating from EBs, where the ingress and egress of the eclipse events could be mis-classified as flare candidates. For this purpose, we use the entire \tess\ threshold-crossing-event (TCE) catalog provided by the SPOC pipeline. We check if the flare peak times are associated with known eclipses. If a flare candidate falls within the eclipse window (approximated as 3 times the TCE duration), we remove the flare from the catalog.
  \item \textit{Variability/Rotation}: This filter removes false positives originating from the peaks of fast-variable and fast-rotating targets. If the target has $\geq 10$ flares, we compute a Lomb-Scargle periodogram of the target light curve and of the affiliated probability time series. If the variability detected from each data sets are within 2 days and have a false-alarm probability $< 0.05$, all flares from this target are removed.
\end{enumerate}

After these filters are applied, we are left with a catalog of $N_{\rm flares} = 958,659$ originating from $N_{\rm stars} = 161,836$ for our analysis. A summary of our catalog is presented in Table~\ref{tab:stats} and the full catalogs are made available in G{\"u}nther et al. (in prep.). We calculate the flare rate per star, $\beta_{\textrm{star}}$, as 
\begin{equation}
  \beta_{\textrm{star}} = 
  \frac{1}{\tau_\textrm{obs}}\,\bigg(\,\sum_{i=1}^{N}p_i\bigg)\,,
\end{equation}
where $N$ is the number of flares for a given star, $p_i$ is the \stella\ probability for a given flare indexed by $i$, and $\tau_\textrm{obs}$ is the total observed time. An overview of the flare rates for all stars in our sample is presented in Figure~\ref{fig:hr}, on a Gaia color-magnitude diagram. Here, we use B$_\textrm{p}$ - R$_\textrm{p}$ as our color and the absolute Gaia G magnitude M$_\textrm{G}$. It is evident that stars with the highest flare rates ($\geq 0.3$\,day$^{-1}$) fall along the upper edge of the main sequence (meaning they are brighter, i.e. have a lower absolute magnitude, than other stars of that given color). This trend could be indicative of binary star systems \citep{hurley98} or young, metal-rich stars \citep{kotoneva02}. For our analysis, we only account for stars along the main sequence and the red giant branch (RGB; stars that fork to the right at 1 $\leq$ Gaia B$_p$-R$_p$ and M$_\textrm{G} \leq 2.5$).

\section{Measured Flare Rates}

\begin{figure}[!ht]
\begin{center}
\includegraphics[width=0.465\textwidth]{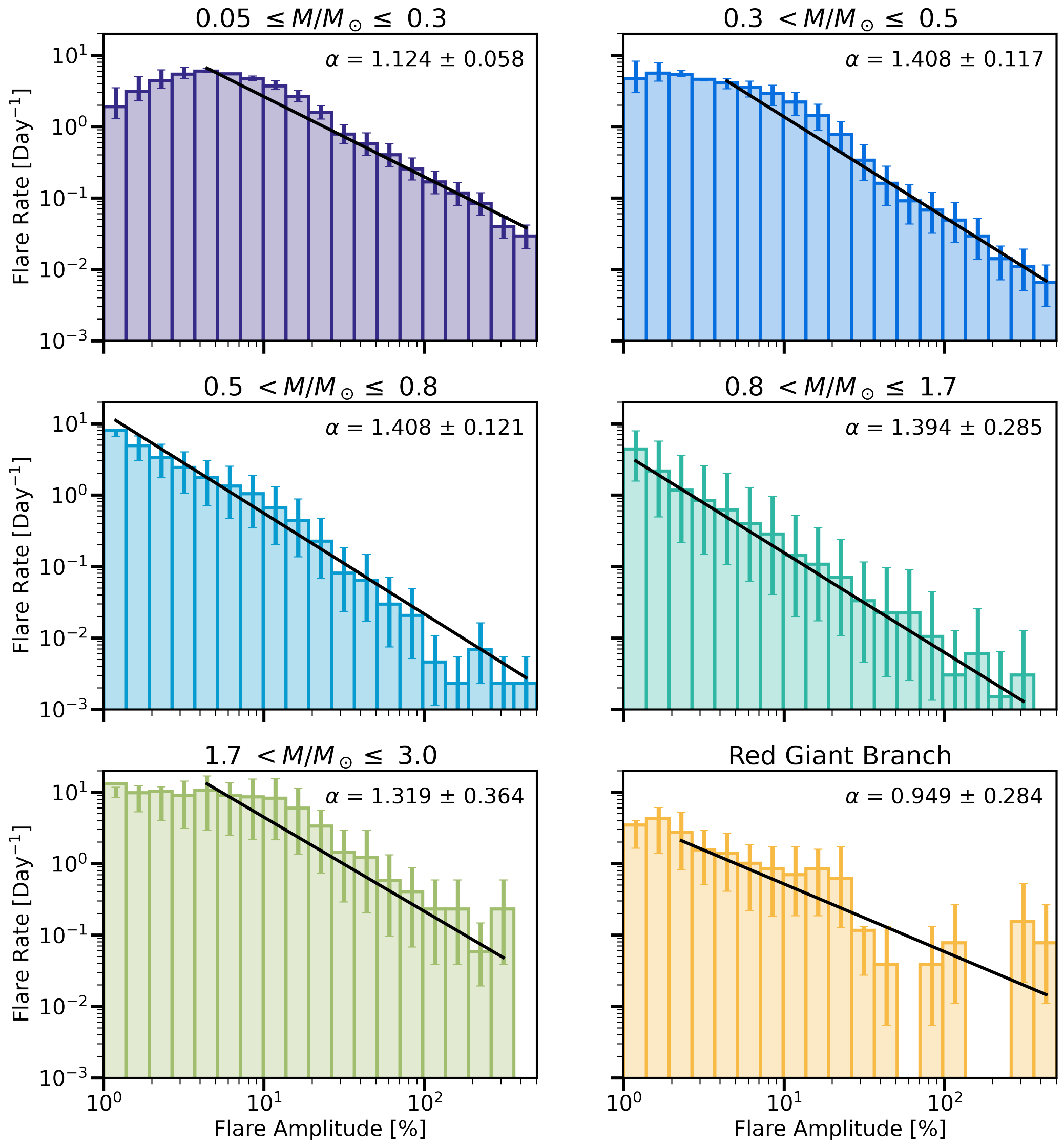}
\caption{The cumulative flare frequency distributions (FFDs) in our sample of stars binned by the flare amplitude and subdivided into different mass bins; the slope, $\alpha$, and error is given in the upper-right corner of each subpanel. The bins are the FFD for flares with a probability $\geq 0.9$. The upper and lower errors on the FFD are defined as flares with probability $\geq 0.99$ and $\geq 0.5$. All bins exhibit clear power-laws, although some bins are incomplete for low-amplitude flares (e.g., $0.05 \leq M/M_\odot \leq 0.3)$ or high-amplitude flares (e.g., Red Giant Branch) . \label{fig:rates}}
\end{center}
\end{figure}

We present our measured flare frequency distributions (FFDs) as a function of stellar mass and flare amplitude (Figure~\ref{fig:rates}). The mass bins were selected based on spectral types/changes in interior stellar structure: stars with masses $M/M_\odot \lesssim 0.3$ are fully convective \citep{Chabrier97, dorman89}, stars with masses $0.3 < M/M_\odot \lesssim 1.7$ have convective exteriors and radiative interiors, and stars with masses $M/M_\odot > 1.7$ have radiative exteriors and convective interiors \citep{heger00}. We present these results in flare amplitude space, compared to energy, to remove any additional errors resulting from estimating stellar luminosities. In theory, the flare energy is more directly relevant to the predicted self-organized critical state. However, in practice, the uncertainty in mass and luminosity of each star makes the amplitude a more reliable quantity. 

\begin{deluxetable}{l r r r}[!ht]
\tabletypesize{\footnotesize}
\tablecaption{Sample Summary Statistics \label{tab:stats}}
\tablehead{\colhead{$B_p - R_p$} & \colhead{Mass (M$_\odot$)} & \colhead{$N_{\textrm{stars}}$} & \colhead{$N_{\textrm{flares, p} \geq 0.5}$}\\
\hline
$[ 2.78 , 4.86 ]$ & $[ 0.05 , 0.3 )$ & 9,241 & 59,150\\
$[ 2.13 , 2.78 )$ & $[ 0.3 , 0.5 )$ & 20,124& 108,963\\
$[ 1.21 , 2.13 )$ & $[ 0.5 , 0.8 )$ & 17,914 & 139,445\\
$[ 0.327 , 1.21 )$ & $[ 0.8 , 1.7 )$ & 85,609 & 571,556\\
$[ -0.12 , 0.327 )$ & $[ 1.7 , 3.0 ]$ & 3,770 & 20,447\\
Red Giant Branch & [0.8, 1.8] & 5,157 & 10,965
}
\startdata
\enddata
\tablecomments{The relationship between Gaia $B_p-R_p$ and stellar mass was taken from \cite{pecaut03}. Red giant branch stellar masses are adopted from \citep{wu19}.}
\end{deluxetable}

We measure the slope of each distribution, $\alpha$, and associated error using a Markov Chain Monte Carlo (MCMC) approach with \texttt{emcee} \citep{Goodman10, Foreman-Mackey13}. Our MCMC chains was initialized with 300 walkers and ran for 5000 steps. After visual inspection, we removed the first 800 burn-in steps and verified our chains converged following the method of \citep{Geweke92}. We present our FFDs and measured slope with error in Figure~\ref{fig:rates}. The black line in each sub-panel represents our fits that excluded incomplete bins. For example, several of the low-amplitude bins are incomplete because of (1) observational constraints, such as the cadence of the observations or stellar/systematic noise in the light curves and (2) limitations imposed by our definition of flare events in \tess\ data (see itemized list of filters in Section~\ref{subsec:id} for more details). Although it is theoretically feasible to quantify the incompleteness via injection-recovery tests, \citet{feinstein20} stated that this method does not accurately represent the performance of the \stella\ CNNs. This limitation is due to  the uncertainty in generating synthetic flare photometric models, and the subsequent differences in real and injected flares. We note that \citet{feinstein20} demonstrated that $\approx 80\%$ of flares with amplitudes $\leq 5\%$ were recovered, corresponding to the first four bins of each panel in Figure~\ref{fig:rates}. For most of our sub-samples, these bins were not used for the fit. Finally, we note that the choice of excluded bins (here due to incompleteness) can affect the power-law slopes resulting from the fits.

\begin{figure}[t]
\begin{center}
\includegraphics[width=0.46\textwidth,trim={0.25cm 0 0 0}]{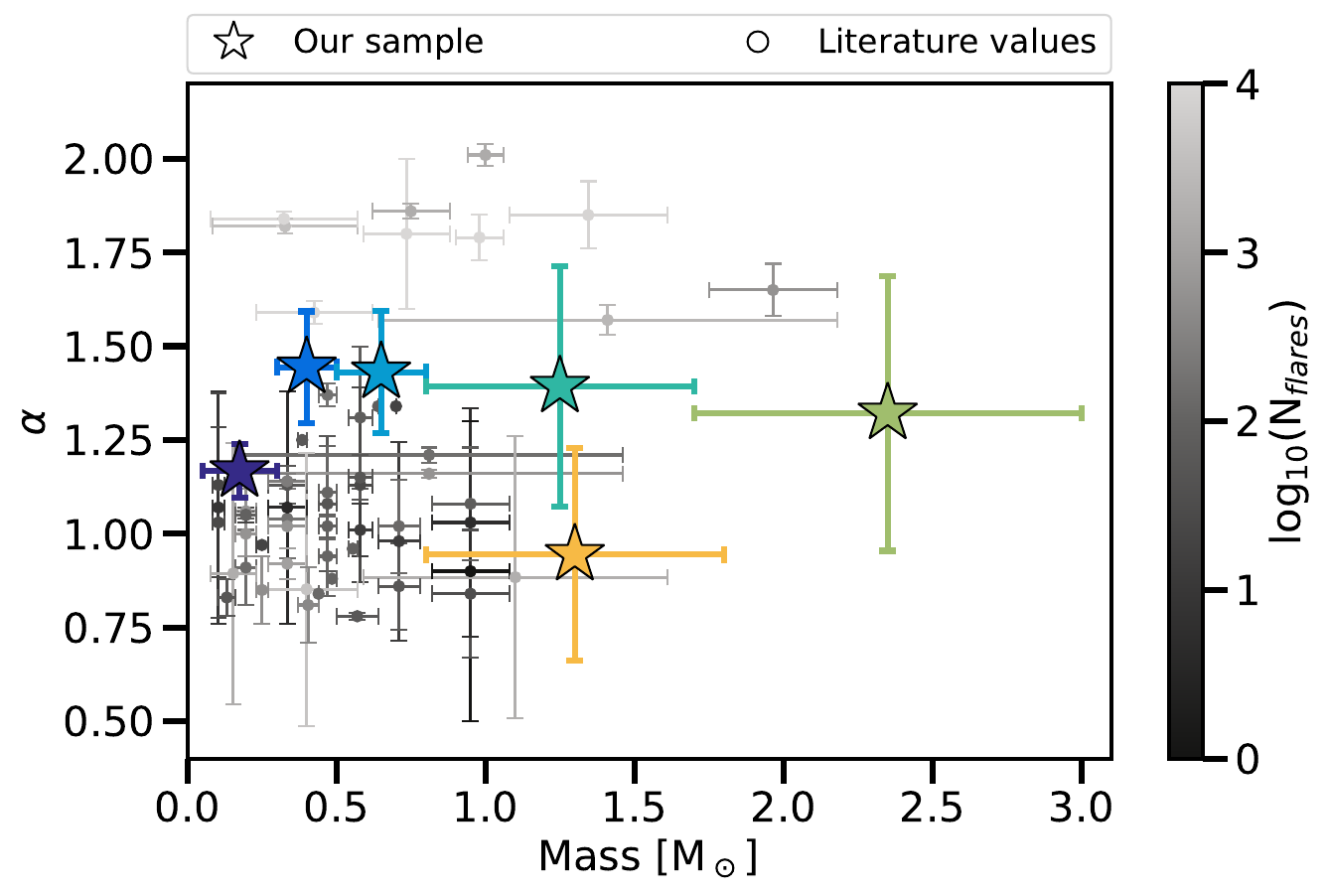}
\caption{Comparison of measured cumulative distribution flare rate slopes, $\alpha$, as a function of stellar mass. Our rates are plotted as stars. Literature values are plotted as circles and colored by the number of flares in the given sample, which range from single to $10^4$ stars \citep{shibayama13, ilin19, lin2019, howard_ward19, yang19, feinstein20, guenther19_flares, raetz20, ilin21, Aschwanden21}. The highest mass stars have higher flare rate indices (light green) than previously measured. Our RGB star (yellow) slope is within $1\sigma$ to that of main sequence stars in the same mass range. The remaining data points fall within the expected range of self-organized critical systems. We estimate the masses of our RGB stars using results from \citep{wu19}. \label{fig:compare}}
\end{center}
\end{figure}

We find that the FFDs for stars with masses $0.3 \leq M/M_\odot \lesssim 3.0$ appear as featureless power-laws. The measured slopes are consistent with models of flaring activity as self-organized critical systems \citep[$\alpha \approx 1.4$][]{lu1991} and with that measured for the Sun  \cite[$\alpha = 1.65\pm0.1$;][]{deArcangelis06}. 

Our lowest mass bin ($0.05 \leq M/M_\odot \leq 0.3$) and our sample of RGB stars do not follow the same trends. Stars with $0.05 \leq M/M_\odot \leq 0.3$ show a featureless power-law for flare amplitudes $\geq 5\%$; while power-laws are indicative of a self-organized critical state, the difference in interior structure may result in a shallower flare rate (here, $\alpha\sim1$). The most energetic flares have historically been observed on low-mass stars in this bin \citep{feinstein20}. It is possible that because these stars are fully convective and therefore have larger convective cells than other main sequence stars, they produce more energetic flares. Similarly, red giants have larger convective cells and display even shallower slopes in their FFD. We also note that our bins of stars with $1.7 < M/M_\odot \leq 3.0$ and RGB stars have the fewest number of flares (Table~\ref{tab:stats}) and while we are able to measure the slopes of these FFDs, they are incomplete to the lowest- and highest-amplitude flares.

Finally, we  compare our measured FFD slopes from our cumulative distribution to those presented by previous authors in Figure~\ref{fig:compare}. Previous studies used an order of magnitude fewer flaring events than ours, and tend to report shallower slopes ($\alpha \sim 1$) than the ones measured here. The variation in measured slopes arises based on sample size/selection, and the cadence of the analyzed data (from 1- to 30-minute cadence). This discrepancy may also be explained by fewer high-amplitude events in the samples, given their rarity. In this study, we also extended our measurement of flare rates to include a broader stellar mass and evolutionary stage range. 

\section{Conclusions \& Future Work}

In this letter, we analyzed the light curves of 161,836 stars that were observed by \tess\ at a 2-minute cadence. These light curves were processed and searched for flares using a novel machine learning algorithm \citep{feinstein20_joss,feinstein20}, resulting in a catalog of $\sim 10^6$ flaring events (G{\"u}nther et al., in prep.). In Figure~\ref{fig:hr}, we show how the flare rate changes as a function of location on a color-magnitude diagram. In Figure~\ref{fig:rates}, we show the flare frequency distribution of stars as a function of stellar mass and compare it to that found in the literature (Figure~\ref{fig:compare}). Main sequence stars with $M/M_\odot \geq 0.3$ exhibit power-law distributions of flare rates with slopes that are characteristic of self-organized critical systems ($\alpha \approx 1.4$). The resulting indices are somewhat smaller for stars with $M/M_\odot < 0.3$ and for red giant stars (which have slopes $\alpha\sim1$). This discrepancy may be due to differences in the interior structure of these stars, which have larger convective cells compared to other main sequence stars, or due to the incompleteness of our sample of flares on red giant branch stars. 

Although the measured slopes of the flare rates show some scatter, the results of this paper indicate a high degree of universality. In the working picture that emerges, subsurface convective regions efficiently inject sufficient energy in the form of twisted magnetic fields near the stellar surface. Subsequent reconnection events then act to produce flares and maintain the magnetic field topology in a self-organized critical state. For completeness, we note that power-law distributions can arise through a variety of mechanisms \citep{newman2005}, not only via self-organized criticality, so that additional theoretical modeling of these systems is indicated. 

\tess\ will continue to observe $\sim 90\%$ of the sky for the next three years. The full-frame images that observe $\sim 10^6$~stars per month will soon increase to a higher cadence, making flare identification more feasible for an order of magnitude more stars than presented herein. This increase should further complete the distribution of the rarest, high-energy events. The timing of sympathetic flares after a ``main" flare event (defined by some amplitude threshold), can be used to further investigate if these systems maintain a self-organized critical state, as was observed for the Sun and for earthquake aftershocks \citep{deArcangelis06}. Due to the short baseline of \tess\ ($\sim 1$\,month), this study may only be truly complete for stars within \tess's continuous viewing zone, those that get $\sim 1$\,year of continuous observations. Such a study could also be completed with the already available \textit{Kepler} light curves, which have a baseline of 4-years to search for the timing of sympathetic flares.

\vspace{10mm}
We thank Mark Krumholz, Fausto Cattaneo, Adrian Price-Whelan, Jacob Bean, Leslie Rogers and Konstantin Batygin for insightful conversations. We thank the anonymous referee and our scientific editor, Manolis K. Georgoulis, for helpful comments and suggestions. ADF acknowledges support from the National Science Foundation Graduate Research Fellowship Program under Grant No. (DGE-1746045). MNG acknowledges support from MIT’s Kavli Institute as a Juan Carlos Torres Fellow and from the European Space Agency  (ESA) as an ESA Research Fellow. This research has made use of NASA's Astrophysics Data System Bibliographic Services. This work made use of the following open-source Python packages \citep{2020NumPy-Array, astropy:2013, astropy18, 2020SciPy-NMeth, matplotlib}.

\bibliography{bibs}
\bibliographystyle{aasjournal}

\end{document}